\documentclass[%
aps,%
showpacs,
nofootinbib,
twocolumn,
superscriptaddress
]{revtex4}
\usepackage{epsfig}
\usepackage{amsmath}

\newcommand{\rmi}[1]{{\mbox{\scriptsize #1}}}

\def\be{\begin{equation}}
\def\ee{\end{equation}}
\def\ba{\begin{eqnarray}}
\def\ea{\end{eqnarray}}
\def\lsi{\raise0.3ex\hbox{$<$\kern-0.75em\raise-1.1ex\hbox{$\sim$}}}
\def\gsi{\raise0.3ex\hbox{$>$\kern-0.75em\raise-1.1ex\hbox{$\sim$}}}
\newcommand{\lsim}{\mathop{\;\lsi\;}}

\begin{document}

\title{Is there a new physics between electroweak and Planck scales?
\footnote{Talks
given at the Workshop on  Astroparticle Physics: Current Issues
(APCI07), Budapest, Hungary, June 21-25, 2007 and at the 11th Paris
Cosmology Colloquium 2007, Paris, France,  August 16-18, 2007.}}

\author{Mikhail Shaposhnikov}
\email{Mikhail.Shaposhnikov@epfl.ch}
\affiliation{
Institut de Th\'eorie des Ph\'enom\`enes Physiques,
Ecole Polytechnique F\'ed\'erale de Lausanne,
CH-1015 Lausanne, Switzerland}

\date{\today}


\begin{abstract}
We argue that there may be no intermediate particle physics energy
scale between the Planck mass $M_{Pl}\sim 10^{19}$ GeV and the
electroweak scale $M_W \sim 100$ GeV. At the same time, the number of
problems of the Standard Model (neutrino masses and oscillations,
dark matter, baryon asymmetry of the Universe, strong CP-problem,
gauge coupling unification, inflation) could find their solution at
$M_{Pl}$ or $M_W$. The crucial experimental predictions of this point
of view are outlined.
\end{abstract}


\maketitle

\section{Introduction}
\label{se:intro}
In this paper we describe a (hopefully) consistent scenario for
physics beyond the Standard Model (SM)  that does not require
introduction of any new energy scale besides already known, namely
the  electroweak and the Planck scales, but can handle different
problems of the SM mentioned in the abstract. This point of view,
supplemented by a requirement of simplicity, has a number of
experimental predictions which can be tested, at least partially,
with the use of existing accelerators and the LHC and with current
and future X-ray/$\gamma$-ray telescopes. Most of the arguments in
favour of the absence of an intermediate energy scale presented in
this work have already appeared in scattered form in refs.
\cite{Asaka:2005an}-\cite{Gorbunov:2007ak}; here they are collected
together with some extra views added.

The paper is organised as follows. In Section \ref{se:extention} we
will review different arguments  telling that the SM model cannot be
a viable effective field theory all the way up to the Planck scale. 
In Section \ref{se:intermediate} we will discuss different arguments
{\em in favour} of existence of the intermediate energy scale and
their weaknesses. In Section \ref{se:numsm} we discuss a proposal for
the physics beyond the SM based on an extention of the SM which  we
called the $\nu$MSM. Section \ref{se:tests} is devoted to the
discussion of crucial tests and experiments that can confirm or rule
out this scenario.

\section{Necessity of extention of the Standard Model}
\label{se:extention}
The content of this section is fairy standard, we need it to fix the
starting point for the discussion that will follow later.

There are no doubts that the Standard Model defined as a
renormalisable field theory based on SU(3)$\times$SU(2)$\times$U(1)
gauge group and containing three fermionic families with left-handed
particles being the SU(2) doublets, right-handed ones being the SU(2)
singlets (no right-handed neutrinos) and one Higgs doublet is not a
final theory. On field-theoretical grounds, it is not consistent as
it contains the U(1) gauge interaction and a self-coupling for the
Higgs field, both suffering  from the triviality, or Landau-pole
problem \cite{Landau:1955ip,Wilson:1973jj,Frohlich:1982tw}. Though
the position of this pole may correspond to experimentally
inaccessible energy scale, this  calls for an ultraviolet (UV)
completion of the theory. 

The existence of gravity with the coupling related to the Planck
scale  $M_{Pl}=G_N^{-1/2}=1.2\times 10^{19}$ GeV ($G_N$ is the
Newtonian gravitational constant) allows to put forward the
hypothesis that the Landau pole problem is solved somehow by a
complete theory that includes the quantum gravity\footnote{In fact,
the scale of quantum gravity may happen to be much smaller than 
$M_{Pl}$. An example is given by theories with large extra dimensions
\cite{ArkaniHamed:1998rs,Antoniadis:1990ew}. We are not going to
consider this possibility here.}. In that way the triviality  is
``swept under the carpet'', provided the position of the Landau pole
is above the Planck scale. It is generally accepted that if the pole
occurs {\em below} the Planck mass then the UV completion of the SM
must not be related to gravity.

It is well known that the requirement that the Landau pole in the
scalar self-coupling  must not appear below some cutoff scale
$\Lambda$ puts an {\em upper bound} on the mass $M_H$ of the Higgs
boson  \cite{Maiani:1977cg,Cabibbo:1979ay,Lindner:1985uk}
(consideration of the U(1) coupling does not produce any extra
information). According to the recent computation
\cite{Pirogov:1998tj}, if $\Lambda$ is identified with the Planck
scale, then $M_H < (161.3\pm 20.6)^{+4}_{-10}$ GeV, where the first
error is theoretical while the second one is related to the
uncertainties in the mass of the top quark. Similar result $M_H <
(180\pm 4)^{+5}_{-5}$ GeV is found in \cite{Hambye:1996wb}. So, a
conservative upper limit is $M_H < 189$ GeV which is well below the
95\% CL limit on the Higgs mass $M_H < 285$ GeV coming from precision
tests of the SM \cite{:2005em}.

In fact, for sufficiently small Higgs masses the SM vacuum is
unstable \cite{Krasnikov:1978pu,Hung:1979dn,Politzer:1978ic}, which 
leads to a {\em lower bound} on the Higgs mass, depending on the
value of the cutoff $\Lambda$, which we take to be $\Lambda=M_{Pl}$.
The computations of this bound done in
\cite{Altarelli:1994rb,Casas:1994qy,Casas:1996aq} lead to $M_H >134
\pm 5$ GeV, consistent with the result of \cite{Pirogov:1998tj} $M_H
>140.7 \pm 10$ GeV. So, a conservative lower limit on the Higgs mass
from these considerations is $M_H >129$ GeV. 

To summarize: theoretically it is possible to think that the SM is
valid all the way up to the Planck scale\footnote{Note that the SM
{\em is not} a consistent field theory up to $M_{Pl}$ if the fourth
chiral family of fermions is introduced \cite{Pirogov:1998tj}.}, and
some complete theory takes over above it, though this is only
feasible if the Higgs mass lies in the interval 
\be
M_H \in [129,189] ~~{\rm  GeV}~.
\label{corr}
\ee 
This region expands at both lower and upper bounds if the scale of
quantum gravity is smaller than the Planck mass.

Let us see now if this point of view can survive when confronted with
different experiments and observations. Since the SM is not
a fundamental theory, the low energy Lagrangian can contain all sorts
of higher-dimensional SU(3)$\times$SU(2)$\times$U(1) invariant 
operators, suppressed by the Planck scale:
\be
L = L_{\rm SM} +\sum_{n=5}^\infty\frac{O_n}{M_{Pl}^{n-4}}~.
\label{lagr}
\ee
These operators lead to a number of physical effects that cannot be
described by the SM, such as neutrino masses and mixings, proton
decay, etc. For example, the lowest order five-dimensional operator
\be
O_5 = A_{\alpha\beta}
\left({\bar L_\alpha}\tilde\phi\right)
\left(\phi^\dagger L_\beta^c\right)
\label{dim5}
\ee
leads to Majorana neutrino masses of the order $m_\nu\sim v^2/M_{Pl}
\simeq 10^{-6}$ eV (here $L_\alpha$ and $\phi$ are the left-handed
leptonic doublets and the Higgs field, $c$ is the sign of charge
conjugation, $\tilde\phi_i=\epsilon_{ij}\phi_j^*$ and $v=175$ GeV is
the vacuum expectation value of the Higgs field). The six-dimensional
operators like $O_6 \propto QQQL$ ($Q$ is the quark doublet) lead to
the proton decay with a life-time exceeding $\tau_p \simeq
M_{Pl}^4/m_p^5 \simeq 10^{45}$ years ($m_p$ is the proton mass).

The fact that $m_\nu$ following from this Lagrangian is so small in
comparison with the lower bound on neutrino mass coming from the
observations of neutrino oscillations $m_\nu > \sqrt{\Delta
m^2_{atm}}=m_{\rmi{atm}}\simeq 0.05$ eV ($\Delta m^2_{atm}$ is the
atmospheric neutrino mass difference, for a review see
\cite{Strumia:2006db}) rules out the conjecture that the SM is a
viable effective field theory up to the Planck scale. Though it is
enough to kill a theory just by one observation, let us discuss
another three, though not so solid ones as they related to
cosmological observations  rather than particle physics experiments. 

(i) Since the SM has no candidate for the Dark Matter (DM)
particle\footnote{Neutrinos in the SM are massless and therefore
cannot play a role of DM. Even adding a mass to them does not save
the situation since they can only be a ``hot'' DM candidate, which is
excluded by the structure formation considerations
\cite{Bond:1980ha}.  Another possible SM dark matter objects, made of
baryons (MACHOs - Massive Astrophysical Compact Halo Objects), are
not allowed by the astrophysical bounds on their abundance
\cite{Alcock:2000ph} and  by the Big-Bang Nucleosynthesis
considerations \cite{Steigman:2005uz}.} and the theory (\ref{lagr})
does not contain any new degrees of freedom, a hope to get the DM may
be associated with the primordial black holes (BH)  (for a review see
\cite{Carr:2005zd}). However, those BH which were formed at
temperatures above $\sim 10^{9}$ GeV should have been evaporated by
now \cite{Hawking:1974sw} (see, however, \cite{Adler:2001vs} and
references therein for discussion of the possibility that BH of
Planck mass could be stable). The production of BH at later times
could be enhanced due to the first order electroweak or QCD phase
transitions \cite{Kapusta:2007dn}. However, both of these transitions
are in fact smooth crossovers (see \cite{Kajantie:1996mn} for the EW
case and \cite{Aoki:2006we} for QCD) and the number of produced BH is
far too small to play a role of DM. So, unless a complete theory
provides the stable states such as  non-evaporating BH or some
particles with the Planck mass, the theory (\ref{lagr}) fails to
describe the dark matter in the Universe.  

(ii) The theory (\ref{lagr}) does not contain any scalar field that
could play a role of the inflaton and thus the inflation should occur
due to some Planck scale physics as was proposed already in 
\cite{Starobinsky:1979ty,Starobinsky:1980te}. However, the vacuum
energy density $V$ during inflation is limited from above
\cite{Rubakov:1982df} by (non) observation of tensor fluctuations of
the cosmic microwave background radiation, with the current limit
being much smaller than the Planck scale, $V_{inf} \lsim 
(10^{-3} M_{Pl})^4$ \cite{Spergel:2006hy}. This difference
seems to make the pure gravitational origin of inflation unlikely and
so for the theory (\ref{lagr}). Interestingly, a possible alternative
to inflation, the pre Big-Bang scenario of
\cite{Veneziano:1991ek,Gasperini:2002bn}, besides the pure
gravitational degrees of freedom contains a new field -- dilaton,
which may be light and is essential for realisation of the scenario. 

(iii) Though the SM has all the ingredients \cite{Sakharov:1967dj} to
produce the baryon asymmetry of the Universe \cite{Kuzmin:1985mm}, it
fails to do so since there is no first order EW phase transition with
experimentally allowed Higgs boson masses \cite{Kajantie:1996mn}. In
addition, it is a challenge to use CP-violation in
Cabibbo--Kobayashi--Maskawa mixing of quarks for net baryon
production. The higher dimensional operators can only contribute to
baryogenesis provided the temperature is of the order of the Planck
scale, whereas the maximum reheating temperature after inflation from
the argument above is at most $10^{16}$ GeV. Thus, the Lagrangian 
(\ref{lagr}) has to be modified.

In addition to experimental and observational drawbacks of the SM one
usually adds to the list of its problems different naturalness
issues, such as: ``Why the EW scale is so much smaller than the
Planck scale?'', ``Why the cosmological constant is so small but
non-zero?'', ``Why CP is conserved in strong interactions?'', ``Why
electron is much lighter than $t$-quark?'' etc., making the 
necessity of physics beyond the SM even more appealing.

\section{Arguments in favour of intermediate energy scale and why they
could be irrelevant}
\label{se:intermediate}
There is the dominating point of view that we must have some new
particle physics between the electroweak scale and the Planck mass.
Let us go through these arguments and try to see whether they are
really convincing. 

{\bf GUT and SUSY scales.} 
We start with gauge coupling unification \cite{Georgi:1974yf}. If one
uses the particle content of the SM and considers the running of the
three gauge couplings one finds that they intersect with each other
at three points scattered between $10^{13}$ and $10^{17}$ GeV (for a
recent review see \cite{Raby:2006sk} ). This is considered as an indication
that strong, weak and electromagnetic interactions are the parts of
the gauge forces of some Grand Unified Theory (GUT) based on a simple
group like SU(5) or SO(10) which is spontaneously broken at energies
$M_{GUT} \sim 10^{16}$ GeV which is close, but still much smaller
than the Planck scale. The fact that the constants do not meet at the
same point is argued to be an indication that there must exist  one
more intermediate threshold for new physics  between the GUT scale
and the electroweak scale, chosen in such a way that all the three
constants do intersect at the same point. The most popular proposal
for the new physics below the GUT scale is the low energy
supersymmetry (SUSY). Indeed, it is amazing that the gauge coupling
unification is almost perfect in the Minimal Supersymmetric Standard
Model (MSSM) \cite{Dimopoulos:1981yj} or in the models based on split
SUSY \cite{ArkaniHamed:2004fb,Giudice:2004tc}. So, these
considerations lead to the prediction of {\em two} intermediate
energy scales between $M_W$ and $M_{Pl}$: one in the potential reach
of the LHC whereas the other can only be revealed experimentally by
the search of proton decay or other processes with baryon number
non-conservation.

The arguments presented above are the standard ones. Before
discussing an alternative let us go through the problem of gauge
hierarchy which exists in the GUT scenario (why and how  achieve
``naturally'' $M_H \ll M_{GUT}$) in some more details. 

Usually the problem of gauge hierarchies is identified with the
problem of quadratic divergences which appear if quantum field theory
is regularized  with the use of a scheme that depends explicitly on
some mass parameter $\Lambda$  (this could be, for example, the UV
cutoff, or Pauli-Villars, or lattice regularizations). One can hear
often that since the quantum corrections to the Higgs scale diverge
quadratically, one must introduce new physics which cancels these
divergences, and that new physics should appear close to the scale
under consideration (EW in our case). This argument, if applied to
other (quartically) divergent quantity of the SM, the vacuum energy
$\epsilon_{vac}$ (cosmological constant), would lead to necessity of
new physics at energies larger than $\epsilon_{vac}^{1/4}\simeq
10^{-3}$ eV. Since this is not observed, we should either conclude
that the case of quartic divergences is very much different from the
case of quadratic divergences, or accept that this type of logic can
be wrong. And, in fact, besides the problem of Landau pole, the EW
theory itself is known to be a perfectly valid theory without any new
physics \cite{'tHooft:1971rn}.

In GUTS, the SM is an effective field theory below $M_{GUT}$, having
a {\em field theory} UV completion above $M_{GUT}$. This makes the
cutoff scale to be a physical parameter (so it must not be sent to
infinity); and to achieve that the {\em physical} Higgs mass is much
smaller than the GUT scale one has to choose carefully  
counter-terms up to $N \simeq
\log(M_{GUT}^2/M_W^2)/\log(\pi/\alpha_W)\simeq 13$ loop level in
non-supersymmetric GUTS \cite{Gildener:1976ai}, which is considered
to be an enormous fine-tuning (and one has to do it $13$ times!) and
as an argument for existing of new physics right above the EW scale.
The low energy SUSY extentions of the SM ease the problem: it is
sufficient to make a fine tuning just once, at the tree level, and
all loop corrections will cancel away automatically, which is
believed to be an advantage of the MSSM with respect to the
SM\footnote{Incidentally, the author of this note does not see much
an advantage of the SUSY here. Admittedly, it is very hard if not
impossible for a {\em theorist} to do the computation of the $13$
loop graphs. However, why one should think in terms of unphysical
quantities  that appear as intermediate steps of perturbative
expansion rather than physical, on-shell ones?}.

In fact, the structure of divergences does depend on the
regularisation scheme. One can use also the  scale-independent
regularisation, such as the dimensional regularisation of 't Hooft
and Veltman \cite{'tHooft:1972fi}.  In this scheme the
renormalization of parameters is the multiplicative one and thus
there is no difference in removing the divergences from dimensionless
parameters such as gauge coupling or dimensionfull parameters such as
the mass of the Higgs boson. However, inspite of these specific
features of the dimensional regularisation the conclusion about fine
tunings remains intact \cite{Weinberg:1980wa}: even in this scheme to
have a {\em field theory GUT}  with two or more well separated scales
one has to tune a number (varying from $1$ in SUSY GUTs to $14$ in
non-SUSY GUTs) of terms to achieve the hierarchy of masses.

Is there an alternative to this logic which removes the necessity of
introduction of these intermediate scales? Perhaps, a simplest
possibility is to say that there is no Grand Unification and the fact
that the gauge couplings nearly meet at high energy scale is a pure
coincidence. Then the ``stand alone'' EW theory contains just one
energy scale, the Higgs mass, which does not require any fine-tuning
if one uses the minimal subtraction scheme of \cite{'tHooft:1972fi}.
True, the theory is not mathematically consistent because of the
Landau pole, but hiding this pole and the vacuum instability above
the Planck scale leaves the solution  for a complete theory of
gravity. Moreover, it is quite possible that the Planck mass cannot
be considered as a {\em field-theoretical} cutoff (or as a  mass of
some particle in the  dimensional regularization) as we still do not
know what happens at the Planck scale. 

Of course, it is a pity to give up the Grand Unification. In addition
to gauge coupling unification GUTs provide an explanation of charge
quantization \cite{Georgi:1974sy}\footnote{In fact, in the SM with
{\em one} fermionic generation the structure of U(1) hypercharges of
all particles can be fixed by the requirement of absence of gauge and
mixed gauge-gravitational  anomalies. The anomaly free solution leads
automatically to charge quantisation \cite{Geng:1988pr}. This is not
the case for the SM with 3 families, where the choice of
hypercharges may be unequal for different generations
\cite{Foot:1990uf}.} and give some non-trivial relations
between quark and lepton masses \cite{Buras:1977yy}. An alternative
is to have gauge coupling unification at the Planck scale. It is
known  \cite{Hill:1983xh,Shafi:1983gz} that this possibility can be
easily realised in GUTs, if higher order non-renormalisable operators
are included in the analysis, 
\be
L = L_{\rm GUT} +\sum_{n=5}^\infty\frac{O_n}{M_{Pl}^{n-4}}~.
\label{gut}
\ee
Indeed, if $F_{\mu\nu}$ is the GUT gauge field strength and $\Phi$ is
the scalar field in adjoint representation which is used to break
spontaneously the GUT group down to the SM, the operators like 
\be
O_{4+n} = {\rm Tr} [F_{\mu\nu}\Phi^k F^{\mu\nu}\Phi^{n-k}]~,~~0\leq
k<n,~n>0
\ee
will rescale the SM gauge couplings with large effect if
$\langle \Phi\rangle \sim M_{Pl}$. It was shown in 
\cite{Parida:1989kg,Brahmachari:1993yn} that it is sufficient to add
dimension $5$ and $6$ operators to the minimal SU(5) theory to bring
the unification scale up to the Planck one. In this case the
corrections due to higher order operators are reasonably small and
within $10$\%. Note that the fact of charge quantization in GUTs does
not depend on the unification scale, while the breaking of the
minimal SU(5) GUT predictions for lightest fermion masses is in fact
welcome. 

To summarize: it is appealing to think that there is no new
field-theoretical scale between $M_W$ and $M_{Pl}$ and that the gauge
couplings meet at $M_{Pl}$ ensuring that {\em all four interactions}
get unified at one and the same scale. This is only self-consistent
if the Higgs mass lies in the interval (\ref{corr}). The experimental
fact that $M_H \ll M_{Pl}$ gets unexplained, but the absence of any
{\em field-theoretical} cutoff below the Planck mass makes this
hierarchy stable, at least in the minimal subtraction renormalization
scheme.

{\bf Inflation.} 
The energy density of the Universe at the exit from inflation
$V_{inf}$ (for a review and historical account of inflationary
cosmology see \cite{Linde:2007fr}) is not known and may vary from $(2
\times 10^{16}~{\rm GeV})^4$ on the high end (limit is coming from the
CMBR observations) down to  $({\rm few~MeV})^4$ (otherwise the
predictions of Big Bang Nucleosynthesis will be spoiled). At the same
time, there are the ``naturalness'' arguments telling that $V_{inf}$
should better be large (for a review see \cite{Lyth:1998xn}), as
otherwise it is difficult to reconcile the necessary number of
$e$-foldings with the amplitude of scalar perturbations. The simplest
quadratic potential 
\be
V(\chi)= \frac{1}{2} m_\chi^2 \chi^2
\ee
fits well the data \cite{Spergel:2006hy} with $m_\chi \sim 10^{13}$
GeV. This fact, and also the closeness of this number to the GUT
scale is often considered as an extra argument in favour of existence
of the high energy scale between $M_W$ and $M_{Pl}$.

It is well known, however, that the CBM constraints the 
inflaton potential (say, in the single field chaotic inflation) only
for the inflaton field of the order of the Planck scale and tells
nothing about the structure  of $V(\chi)$  near its minimum (for a
recent discussion see \cite{Lesgourgues:2007gp}). In other words, the
inflaton may be very light whereas the requisite large $V_{inf}$ may
come from its self-interactions. For example, even a pure $\beta
\chi^4$ potential ({\em massless} inflaton) provides a reasonable fit
to the WMAP data with just $3\sigma$ off the central values for
inflationary parameters  \cite{Spergel:2006hy}, which can be
corrected by a slight modification of the potential at $\chi \sim
M_{Pl}$ by higher dimensional operators or by allowing non-minimal
coupling of the inflaton to gravity
\cite{Hwang:1998mx,Komatsu:1999mt}. We will discuss a concrete
proposal for light inflaton interactions later in Section
\ref{se:numsm}.

{\bf Strong CP-problem.}
One of the fine-tuning problems of the SM is related to complicated
vacuum structure of QCD leading to the existence of the vacuum
$\theta$ angle \cite{Jackiw:1976pf,Callan:1976je} leading to
CP-non-conservation in strong interactions. A most popular solution to
the problem is related to Peccei-Quinn symmetry \cite{Peccei:1977hh}
which brings $\theta$ to zero in a dynamical way; a degree of freedom
which is responsible for this is a new hypothetical (pseudo) scalar
particle - axion \cite{Weinberg:1977ma,Wilczek:1977pj} or invisible
axion \cite{Shifman:1979if,Zhitnitsky:1980tq,Dine:1981rt}. Axion has
never been seen yet, and the strong limits on its mass and couplings
are coming from direct experiments \cite{Asztalos:2006kz} and from
cosmology and astrophysics \cite{Raffelt:1990yz}. They lead to an
admitted ``window'' for the Peccei-Quinn scale $10^{8}$ GeV $\lsi
M_{PQ}\lsi 10^{12}$ GeV where the lower and upper bound depend on the
type of axion and different cosmological assumptions. So, it looks
like an intermediate scale appears again!

In fact, the axion solution to the strong CP-problem is not the
unique one. As an example we will discuss shortly a proposal for a
solution of the strong CP-problem which uses extra
dimensions\footnote{Higher spacial dimensions appear, for instance,
in string theory and in  Kaluza-Klein or brane models.} 
\cite{Khlebnikov:1987zg,Khlebnikov:2004am} and does not require the
presence of any new scale between $M_W$ and $M_{Pl}$ and does not
contradict to any observation. Other extra-dimensional solutions have
been suggested in \cite{Aldazabal:2002py,Bars:2006dy}.

The mere existence of the strong CP problem is based on the
assumption that the number of dimensions of the space-time is four.
Indeed, the existence of $\theta$ vacua is related to topology: the
mapping of the three-dimensional sphere, representing our space, to
the gauge group SU(3) of QCD is non-trivial, $\pi_3(S_3)$ = $Z$. This
leads to the existence of classical vacua with different topological
numbers, and the quantum tunneling between these states forms a
continuum of stable vacua characterized by $\theta\in[0,2\pi)$.
Clearly, these considerations are only valid if the space is
$3$-dimensional. Thus, in higher dimensional theories, where the
3-dimensional character of the space is just a low-energy
approximation, the strong CP-problem has to be reanalyzed. 

If the topology of the higher-dimensional space is such that the
mapping of it to the gauge group is trivial, strong CP-problem
disappears. Concrete examples were given
\cite{Khlebnikov:1987zg,Khlebnikov:2004am} for $4+1$ dimensional
space-time, where the space is a 4-sphere $S_4$. It was shown there
that the only remnant from extra dimensions which should be added to
the low energy effective theory is a quantum-mechanical degree of
freedom - ``global axion'' $a(t)$ which depends on time but does not
depend on the spacial coordinates and thus does not represent any new
{\em particle} degree of freedom. The global axion couples to the
ordinary fields in a way the standard axion does and thus relaxes the
effective vacuum $\theta$ angle to zero. Still, the solution of the
U(1) problem in QCD is unaffected \cite{Khlebnikov:2006yq}. None of 
the astrophysical bounds can be applied to the global axion, simply
because there is no particle to emit or absorb, while it is
impossible to excite the $a(t)$ by any local process. As for the
``over-closure of the Universe'' constraint, it depends strongly on
the cosmological scenario of dynamics of the compactification which
may happen at the Planck scale. 

We see therefore that the strong CP-problem, if fact, does not point
to the existence of an intermediate scale.

{\bf Neutrino masses.} 
A popular argument in favour of existence of the very large mass
scale is related to neutrino masses \cite{Seesaw}. Indeed, let us add
to the Lagrangian of the Standard Model  a dimension five operator
(\ref{dim5}) suppressed by an (unknown a-priory) mass parameter
$\Lambda$ and find it then from the requirement that this term gives
the correct active neutrino masses. One gets immediately that
\be
\Lambda \simeq \frac{v^2}{m_{\rmi{atm}}}\simeq 6\times 10^{14}~{\rm
GeV}~,
\label{nuM}
\ee
which is amazingly close to the GUT scale. 

In fact, eq. (\ref{nuM}) provides {\em an upper bound} on the scale
of new physics beyond the SM rather than {\em an estimate} of this
scale. This will be discussed in more detail in  Section 
\ref{se:numsm}.

{\bf Baryogenesis.}
One of the key points of any baryogenesis scenario is departure from
thermal equilibrium \cite{Sakharov:1967dj}. One of the popular
mechanisms is called thermal leptogenesis \cite{Fukugita:1986hr}. In
this scenario heavy Majorana neutrinos $N$ with the mass $M_N$ decay
with non-conservation of lepton number and CP and produce lepton
asymmetry of the Universe which is then converted to baryon asymmetry
in rapid EW anomalous processes  with fermion number non-conservation
\cite{Kuzmin:1985mm}. A very quick (and missing many details)
estimate  shows that $M_N$ should be better close to the GUT scale.
Indeed, the temperature at which $N$ decay should be smaller than
their mass (out of equilibrium condition) but larger than the EW
scale (sphalerons must be active), $M_W < T_{decay} < M_N$. This
leads to the following constraint on the typical Yukawa coupling of
$N$ to the leptons and the SM Higgs (the decay rate of $N$ is
roughly  $\Gamma_{\rm tot} \simeq f^2 M_N$):
\be
\frac{M_W^2}{M_N M_0} < f^2 < \frac{M_N}{M_0}~,
\ee
where  $M_0 \simeq 10^{18}~{\rm GeV}$, and  the Hubble 
constant-temperature  relation is $H=T^2/M_0$.  CP-violating effects
appear from loop corrections to the decay amplitudes, and without
extra fine-tunings one gets an estimate for baryon-to-entropy ratio
\be
\frac{n_B}{s} \sim 10^{-3} f^2 ~.
\ee
A correct prediction is obtained for $f^2 \sim 10^{-7}$, leading to
the requirement $M_N > f^2 M_0 \simeq 10^{11}$ GeV. At the same time,
since temperature after inflation cannot exceed $10^{16}$ GeV, the
leptogenesis with Planck mass Majorana leptons does not seem to be
possible\footnote{Even stronger constraints exist in supergravity
theories coming from the copious gravitino production,  see
\cite{Rychkov:2007uq} for a recent discussion.}.

Electroweak baryogenesis, in which the only source for baryon number
non-conservation is the electroweak anomaly, requires strongly first
order phase transition
\cite{Shaposhnikov:1986jp,Shaposhnikov:1987tw}. As this phase
transition is absent in the SM \cite{Kajantie:1996mn}, the use of EW
anomaly for baryogenesis calls for modification of the scalar sector
of the EW theory by introducing new scalar singlets or doublets and
thus to a new physics in the vicinity of the EW scale.

Though both of these arguments are certainly true for specific
mechanisms of baryogenesis, they are not universal. We will discuss
in more detail in Section \ref{se:numsm} how they are avoided in the
$\nu$MSM.

{\bf Dark matter.}
A particle physics candidate for dark matter must be a long-lived or
stable particle. The most popular  candidates  are related to
supersymmetry (neutralino etc.) or to the axion, which we have
already discussed. The scenario for WIMPs assumes that initially
these particles were in thermal equilibrium and then annihilated into
the particles of the SM. Quite amazingly, if the cross-section of the
annihilation is of the order of the typical weak cross-section (for a
review see \cite{Bertone:2004pz}) one gets roughly correct abundance
of dark matter, suggesting that the mass of DM particles is likely to
be of the order of the EW scale, as it happens, for example, in the
MSSM, and thus to a new physics nearby.

This argument is based on the specific processes by which the dark
matter can be created and destroyed and thus is not valid in general.
In the next section we will discuss the $\nu$MSM dark matter
candidate with completely different properties.

\section{The $\nu$MSM as an alternative}
\label{se:numsm}
In Section \ref{se:extention} we reviewed the arguments that the SM
must necessarily be extended while in Section \ref{se:intermediate}
we argued that the solutions to the problem of gauge coupling
unification and strong CP problem can be shifted up to the Planck
scale. This cannot be done with neutrino masses, and unlikely to be
possible with other problems we have discussed, namely with dark
matter, baryon asymmetry of the Universe and inflation. In this
section we review how a minimal extention of the SM, the $\nu$MSM,
can solve all of them. 

Let us add to the SM three\footnote{Any number of singlet fermions
can be added without spoiling the consistence of the theory. 
However, the number $3$ is the minimal one which allows simultaneous
explanation of neutrino masses and mixings, dark matter and baryon
asymmetry of the Universe, see below. Since the maximal number of
fermionic generations in a theory without intermediate energy scale
is also $3$, it looks reasonable to pick up the same for the singlet
fermions.} right-handed fermions $N_I$, $I=1,2,3$ (they can be called
singlet leptons, right-handed leptons or Majorana neutrinos) and
write the most general renormalisable interaction between these
particles and fields of the SM: 
\ba
L_{\nu MSM}=L_{SM}+
\bar N_I i \partial_\mu \gamma^\mu N_I-\nonumber\\
  - F_{\alpha I} \,  \bar L_\alpha N_I \tilde \phi 
  - \frac{M_I}{2} \; \bar {N_I^c} N_I + h.c.~.
  \label{lagr1}
  \ea
Here $L_{SM}$ is the Lagrangian of the SM, $\alpha=e,\mu,\tau$, and
both Dirac ($M^D = F_{\alpha I} \langle \phi \rangle$) and Majorana
($M_I$) masses for singlet fermions are introduced. This Lagrangian
contains 18 new parameters in comparison with the SM.

Why this Lagrangian? Since we even do not know where the SM
itself is coming from, the answer to this question can only be very
vague. Here is an argument in its favour. The particle content of the
SM has an asymmetry between quarks and leptons: every left quark and
charged lepton has its counterpart - right quark or right-handed
lepton, while the right-handed counterpart for neutrino is missing.
The Lagrangian (\ref{lagr1}) simply restores the symmetry between
quarks and leptons. Interestingly, the requirement of gauge and
gravity anomaly cancellation, applied to this theory, leads to
quantization of electric charges for three fermionic generations
\cite{Foot:1990uf}, which was not the case for the SM, because of new
relations coming from Yukawa couplings and Majorana masses.

Besides fixing the Lagrangian, one should specify the masses and
couplings of singlet fermions. The see-saw \cite{Seesaw} logic picks
up the Yukawa term in (\ref{lagr1}) and tells that it is ``natural''
to have Yukawa coupling constants of new leptons of the same order of
magnitude as Yukawa couplings of quarks or charged leptons. Then the
mass parameters for singlet fermions must be large, $M \sim
10^8-10^{14}$ GeV, to give  the correct order of magnitude for active
neutrino masses. This leads to an intermediate energy scale already
discussed above. Yet another proposal is to fix the masses of singlet
fermions in the eV region \cite{deGouvea:2005er} to explain the LSND
anomaly \cite{Aguilar:2001ty}. Note that the oscillation explanation
of the LSND result is disfavoured by the MiniBooNE experiment
\cite{AguilarArevalo:2007it}.

The $\nu$MSM logic picks up the mass term in (\ref{lagr1}) and
assumes that it is ``natural'' to have it {\em roughly} of the order
of another mass term in the EW Lagrangian, namely that of the Higgs
boson \footnote{We say ``roughly'' since even if the source of the
mass is known and unique, as in the SM case, the numerical values of
particle masses can be very much different. For example, the 
electron is lighter than the top quark by 5 orders of magnitude.}.
This does not lead to any intermediate scale but requires smaller
Yukawa couplings.   To get a more precise idea about the values of
Majorana masses, a phenomenological input, discussed below, is
needed.

{\bf Neutrino masses and oscillations.}
The Lagrangian  (\ref{lagr1}) can explain any pattern of active
neutrino masses and their mixing angles for arbitrary (and, in
particular, below the EW scale) choice of the Majorana neutrino
masses. This is a simple consequence of the parameter counting: the
active neutrino mass matrix can be completely described by $9$
parameters whereas (\ref{lagr1}) contains $18$ arbitrary masses and
couplings.

{\bf Dark matter.}
The dark matter candidate of the $\nu$MSM is the long-lived lightest
singlet fermion. The mass of this particle is not fixed by
theoretical considerations. However, there are some cosmological and
astrophysical arguments giving a preference to the keV region. In
particular, the keV scale is favoured by the cosmological
considerations of the production of dark matter due to transitions
between active and sterile neutrinos \cite{Dodelson:1993je} and by
the  structure formation arguments related  to the problems of
missing satellites and cuspy profiles in the Cold Dark Matter (CDM)
cosmological models
\cite{Moore:1999nt,Bode:2000gq,Goerdt:2006rw,Gilmore:2007fy} (see,
however, \cite{Penarrubia:2007zz}); warm DM may help to solve  the
problem of galactic angular momentum~\cite{Sommer-Larsen:1999jx}. At
the same time, much larger masses are perfectly allowed 
\cite{Shi:1998km,Shaposhnikov:2006xi}; in this case the dark matter
sterile neutrino is a CDM candidate. This particle has never been in
thermal equilibrium in the early Universe and thus the arguments
about the mass scales of the dark matter particle of the previous
section do not apply to it. For  reviews of different  astrophysical
constraints on the properties of the sterile neutrino dark matter,
and the mechanisms of its cosmological production see
\cite{Shaposhnikov:2007nf,Ruchayskiy:2007pq} and references therein.
We just mention here that the {\em simultaneous explanation} of
neutrino masses and mixings and dark matter requires that the number
of singlet fermions in the $\nu$MSM is at least three and that the
mass of one of the active neutrinos is very small, $\lsi ~{\cal
O}(10^{-5})$ eV \cite{Asaka:2005an,Boyarsky:2006jm}. 

{\bf Baryogenesis.}
The phase structure of the $\nu$MSM is the same as that of the SM: 
there is no EW phase transition which could lead to large
deviations from thermal equilibrium. The masses of singlet fermions
are smaller than the electroweak scale, they decay below the
sphaleron freezout temperature and thus the thermal leptogenesis
of \cite{Fukugita:1986hr} does not work. However, the presence of
singlet fermions provides another source of thermal non-equilibrium,
simply because these particle, due to their small Yukawa couplings,
interact very weakly. The mechanism of baryogenesis in this case is
related to coherent resonant oscillations of singlet fermions  
\cite{Akhmedov:1998qx,Asaka:2005pn}. To explain simultaneously
neutrino masses, dark matter and baryon asymmetry of the Universe at
least three singlet fermions are needed, with two of them with the
mass preferably in the GeV region \cite{Shaposhnikov:2006nn}. They
are required to be almost degenerated
\cite{Akhmedov:1998qx,Asaka:2005pn}. The specific pattern of the
singlet lepton masses and couplings leading to phenomenological
success of the $\nu$MSM can be a consequence of the leptonic U(1)
symmetry discussed in \cite{Shaposhnikov:2006nn}.

{\bf Inflation.}
Adding just new fermions to the SM cannot lead to inflation. The
simplest way out is to introduce a singlet scalar field, $\chi$, with
scalar potential 
\be
V(\chi)=\frac{1}{2} m_\chi^2\chi^2 + \lambda_3\chi^3 +\lambda_4\chi^4 
\ee
and renormalisable couplings to the field of the $\nu$MSM.  The
number of these interactions is in fact not that large, possible
terms are those of interaction of $\chi$ with the Higgs field and
singlet fermions,
\be
L_\chi= h_1 \chi^2 \phi^\dagger\phi  + h_2\chi \phi^\dagger\phi +
f_{IJ}\bar {N_I^c} N_J\chi~.
\label{lowen}
\ee
As we have already discussed in Section \ref{se:intermediate}, the
mass of the inflaton can be small, and taking it to be below the
electroweak scale does not contradict to any principles or
observations. Note also that (\ref{lowen}) for large values of
$\chi\sim M_{Pl}$ can be modified due to Planck scale corrections. It
is not difficult to find constraints on the inflaton couplings to the
fields of the $\nu$MSM which insure that the model with inflaton
satisfies experimental, astrophysical, and cosmological constraints
\cite{Shaposhnikov:2006xi}. 

In fact, the number of the parameters in the $\nu$MSM with the
inflaton field can be reduced greatly without loosing its attractive
phenomenological features \cite{Shaposhnikov:2006xi} if one requires
that the classical Lagrangian obeys an (approximative) dilatation
symmetry \cite{Buchmuller:1990pz}. A requirement that the complete
Lagrangian exhibits dilatation symmetry on classical level puts all
dimensional couplings of the theory (mass of the Higgs boson,
inflaton, and masses of singlet fermions, and also $\lambda_3$ and
$h_2$) to zero; in this case the origin of all masses in the $\nu$MSM
must be related to the vacuum expectation value of the inflaton field
$\chi$. This (Coleman-Weinberg \cite{Coleman:1973jx}) scenario can
only work if $\chi$ gets a non-trivial potential due to radiative
corrections. Though possible for a specific choice of  $\lambda_4$
and $f_{IJ}$ \cite{Meissner:2006zh}, the theory obtained cannot
accommodate inflation (since $\lambda_4$ is required to be of the
order of one \cite{Meissner:2006zh} and thus too large density
perturbations are generated) and baryon asymmetry of the Universe
(since $f_{IJ}$ must be of the order of one \cite{Meissner:2006zh}
which leads these particles to thermal equilibrium well above the
electroweak scale), and, therefore, the requirement of {\em complete}
dilatation invariance should be abandoned. A way out is to break it
by minimal means, and a possibility is to admit that $m_\chi^2 \neq
0$ and negative \cite{Shaposhnikov:2006xi}. For a theory constructed
in such a way the condensate of $\chi$ gives masses to singlet
fermions and induces the EW symmetry breaking; the same field gives
rise to inflation. Moreover, the mass of $\chi$ {\em must} be smaller
than the EW scale: if this is not the case the energy stored in the
inflaton field right after inflation will go predominantly to the
inflaton field itself rather than to the Higgs field and eventually
to other degrees of freedom of the SM. This would change the standard
Big Bang scenario right above the EW scale and make baryogenesis
impossible. In fact, to keep baryogenesis in place a stronger
constraint $m_\chi < 2 M_N$ must be satisfied 
\cite{Shaposhnikov:2006xi}.

{\bf Fine tunings and hierarchies.}
The ``stand alone'' $\nu$MSM or an extention of this model by a light
inflaton is a theory with just one energy scale and thus it does not
suffer from a fine-tuning problem typical to the {\em field theory}
models containing two or more very distinct energy scales. Moreover,
the masses of the singlet leptons are protected by the lepton number
symmetry and thus can ``naturally''  be small. In addition, due to
the smallness of all extra constants of interaction  the
renormalisation group behaviour of the SM couplings remains
practically the same, and  the interval (\ref{corr}) does not change.
Of course, this theory, as the SM, has a Landau pole problem, but it
can presumably be avoided in a more complete theory that includes
gravity, if this pole is situated above the Planck scale. 

\section{Crucial tests and experiments}
\label{se:tests}
As we argued, none of the arguments in favour of existence of the
intermediate energy scale really requires it: gauge coupling 
unification and solution of the strong CP-problem can both occur at
the Planck scale, whereas inflation, neutrino masses, dark matter and
baryogenesis can all be explained by the particles with the masses
below the electroweak scale.

The point of view that there is no intermediate energy scale between
the weak and Planck scales and that the low energy effective theory
is the $\nu$MSM which explains neutrino oscillations, dark matter and
baryon asymmetry of the Universe is rather fragile.  It predicts an
outcome of a number of experiments, and if {\em any} of the
predictions is not satisfied, this conjecture will be ruled out.
Though most of these predictions were discussed elsewhere, we
present them here for completeness.\\
{\bf LHC physics.} Nothing but the Higgs with the mass in the window 
$M_H \in [129,189] ~~{\rm  GeV}$ in which the $\nu$MSM is a
consistent theory below the Planck scale. \\
{\bf ILC physics.} No new physics at the ILC.\\
{\bf Neutrino physics.} Hierarchical structure of active neutrino
masses with one of them smaller than ${\cal O}(10^{-5})$ eV
\cite{Asaka:2005an,Boyarsky:2006jm}. Two other masses are fixed to be
$m_3 = [4.8^{+0.6}_{-0.5}] \cdot 10^{-2}$ eV and $m_2=
[9.05^{+0.2}_{-0.1}]\cdot 10^{-3}$eV ($[4.7^{+0.6}_{-0.5}]\cdot
10^{-2}$eV) in the normal (inverted) hierarchy. Majorana mass of
electron neutrino is smaller than the atmospheric mass difference,
$m_{ee} \leq 0.05$ eV \cite{Bezrukov:2005mx}. The $\nu$MSM is in
conflict with the oscillation hypothesis of the LSND result and with
the result of \cite{Klapdor-Kleingrothaus:2006ff} claiming that the
neutrinoless double $\beta-$ decay has been observed.\\
{\bf Dark matter searches.} Negative result for the WIMP and axion
searches. The existence of a narrow X-ray line due to two-body decays
of the sterile dark matter neutrino. The position and the intensity
of this line are quite uncertain, with a possible cosmological
preference for a few keV energy range, though higher values are
certainly allowed as well. The best choice of astrophysical objects
to search for dark matter sterile neutrino is discussed in
\cite{Boyarsky:2006fg}, and future experimental perspectives in 
\cite{Boyarsky:2006hr}. The laboratory searches of the dark matter
sterile neutrino would require a precision study of kinematics of
$\beta-$decays of tritium or other isotopes \cite{Bezrukov:2006cy}.\\
{\bf B-non-conservation.} No sign of proton decay or
neutron-antineutron oscillations.\\
{\bf Flavour physics.} Existence of two almost degenerate weakly
coupled singlet leptons which can be searched for in rare decays of
mesons or $\tau$-lepton and their own decays can be looked for in
dedicated experiments discussed in \cite{Gorbunov:2007ak}. Though the
masses of these particles cannot be precisely fixed, they must certainly
be below $M_W$  with a preference for small masses $\lsi 1$ GeV
\cite{Shaposhnikov:2006nn}. Visible lepton number non-conservation in
$N$ decays, with CP-breaking that can allow to fix theoretically the
sign and magnitude of the baryon asymmetry of the Universe. Possible
existence of the light inflaton \cite{Shaposhnikov:2006xi}.

\section{Acknoledgements}
\label{se:acknoledgements}
I thank Fedor Bezrukov, Alexey Boyarsky, Sergei Khlebnikov and  Oleg
Ruchayskiy for many helpful comments. This work was supported in part
by the Swiss National Science Foundation.


\begin{thebibliography}{99}

\bibitem{Asaka:2005an}
  T.~Asaka, S.~Blanchet and M.~Shaposhnikov,
  Phys.\ Lett.\ B {631} (2005) 151 
  [hep-ph/0503065].

\bibitem{Asaka:2005pn}
  T.~Asaka and M.~Shaposhnikov,
  Phys.\ Lett.\ B {620} (2005) 17
  [hep-ph/0505013];
    
\bibitem{Shaposhnikov:2006xi}
  M.~Shaposhnikov and I.~Tkachev,
  Phys.\ Lett.\  B {\bf 639} (2006) 414
  [arXiv:hep-ph/0604236].
   
\bibitem{Shaposhnikov:2006nn}
  M.~Shaposhnikov,
  Nucl.\ Phys.\ B {763} (2007) 49
  [hep-ph/0605047].

\bibitem{Gorbunov:2007ak}
  D.~Gorbunov and M.~Shaposhnikov,
  arXiv:0705.1729 [hep-ph].

\bibitem{Landau:1955ip}
L.~D.~Landau, A.~A.~Abrikosov and I.~M.~Khalatnikov,
  Dokl.\ Akad.\ Nauk Ser.\ Fiz.\  {\bf 95} (1954) 773;
{\bf 95} (1954) 1177;{\bf 96} (1954) 261;
L.~D.~Landau and I.~Y.~Pomeranchuk,
  Dokl.\ Akad.\ Nauk Ser.\ Fiz.\  {\bf 102} (1955) 489;
L.~D.~Landau, {\em Niels Bohr and the development of physics}, 
ed. W. Pauli, Pergamon Press, 1955 

\bibitem{Wilson:1973jj}
  K.~G.~Wilson and J.~B.~Kogut,
  Phys.\ Rept.\  {\bf 12} (1974) 75.
  
\bibitem{Frohlich:1982tw}
  J.~Frohlich,
  Nucl.\ Phys.\  B {\bf 200} (1982) 281.
  
  
  
\bibitem{ArkaniHamed:1998rs}
  N.~Arkani-Hamed, S.~Dimopoulos and G.~R.~Dvali,
  Phys.\ Lett.\  B {\bf 429} (1998) 263
  [arXiv:hep-ph/9803315].
  
  
\bibitem{Antoniadis:1990ew}
  I.~Antoniadis,
  Phys.\ Lett.\  B {\bf 246} (1990) 377.
  
\bibitem{Maiani:1977cg}
  L.~Maiani, G.~Parisi and R.~Petronzio,
  Nucl.\ Phys.\  B {\bf 136} (1978) 115.
  
\bibitem{Cabibbo:1979ay}
  N.~Cabibbo, L.~Maiani, G.~Parisi and R.~Petronzio,
  Nucl.\ Phys.\  B {\bf 158} (1979) 295.
   
\bibitem{Lindner:1985uk}
  M.~Lindner,
  Z.\ Phys.\  C {\bf 31}, 295 (1986).
  
\bibitem{Pirogov:1998tj}
  Yu.~F.~Pirogov and O.~V.~Zenin,
  Eur.\ Phys.\ J.\  C {\bf 10} (1999) 629
  [arXiv:hep-ph/9808396].
  
\bibitem{Hambye:1996wb}
  T.~Hambye and K.~Riesselmann,
  Phys.\ Rev.\  D {\bf 55} (1997) 7255
  [arXiv:hep-ph/9610272].
  
  
\bibitem{:2005em}
    [ALEPH Collaboration],
  Phys.\ Rept.\  {\bf 427} (2006) 257
  [arXiv:hep-ex/0509008].
   
\bibitem{Krasnikov:1978pu}
  N.~V.~Krasnikov,
  Yad.\ Fiz.\  {\bf 28} (1978) 549.
  
\bibitem{Hung:1979dn}
  P.~Q.~Hung,
  Phys.\ Rev.\ Lett.\  {\bf 42}, 873 (1979).
  
\bibitem{Politzer:1978ic}
  H.~D.~Politzer and S.~Wolfram,
  Phys.\ Lett.\  B {\bf 82}, 242 (1979)
  [Erratum-ibid.\  {\bf 83B}, 421 (1979)].
  
\bibitem{Altarelli:1994rb}
  G.~Altarelli and G.~Isidori,
  Phys.\ Lett.\  B {\bf 337}, 141 (1994).
  
\bibitem{Casas:1994qy}
  J.~A.~Casas, J.~R.~Espinosa and M.~Quiros,
  Phys.\ Lett.\  B {\bf 342}, 171 (1995)
  [arXiv:hep-ph/9409458].

\bibitem{Casas:1996aq}
  J.~A.~Casas, J.~R.~Espinosa and M.~Quiros,
  Phys.\ Lett.\  B {\bf 382}, 374 (1996)
  [arXiv:hep-ph/9603227].
   
\bibitem{Strumia:2006db}
  A.~Strumia and F.~Vissani,
  arXiv:hep-ph/0606054.
  
  
\bibitem{Bond:1980ha}
  J.~R.~Bond, G.~Efstathiou and J.~Silk,
  Phys.\ Rev.\ Lett.\  {\bf 45} (1980) 1980.
  
\bibitem{Alcock:2000ph}
  C.~Alcock {\it et al.}  [MACHO Collaboration],
  Astrophys.\ J.\  {\bf 542} (2000) 281
  [arXiv:astro-ph/0001272].
  
\bibitem{Steigman:2005uz}
  G.~Steigman,
  Int.\ J.\ Mod.\ Phys.\  E {\bf 15} (2006) 1
  [arXiv:astro-ph/0511534].
  
\bibitem{Carr:2005zd}
  B.~J.~Carr,
  arXiv:astro-ph/0511743.
  
\bibitem{Hawking:1974sw}
  S.~W.~Hawking,
  Commun.\ Math.\ Phys.\  {\bf 43} (1975) 199
  [Erratum-ibid.\  {\bf 46} (1976) 206].
  
\bibitem{Adler:2001vs}
  R.~J.~Adler, P.~Chen and D.~I.~Santiago,
  Gen.\ Rel.\ Grav.\  {\bf 33} (2001) 2101
  [arXiv:gr-qc/0106080].
  
\bibitem{Kapusta:2007dn}
  J.~I.~Kapusta and T.~Springer,
  arXiv:0706.1111 [astro-ph].
  
\bibitem{Kajantie:1996mn}
  K.~Kajantie, M.~Laine, K.~Rummukainen and M.~E.~Shaposhnikov,
  Phys.\ Rev.\ Lett.\  {\bf 77} (1996) 2887
  [arXiv:hep-ph/9605288].
  
  
\bibitem{Aoki:2006we}
  Y.~Aoki, G.~Endrodi, Z.~Fodor, S.~D.~Katz and K.~K.~Szabo,
  Nature {\bf 443}, 675 (2006)
  [arXiv:hep-lat/0611014].
 

\bibitem{Starobinsky:1979ty}
  A.~A.~Starobinsky,
  JETP Lett.\  {\bf 30}, 682 (1979)
  [Pisma Zh.\ Eksp.\ Teor.\ Fiz.\  {\bf 30}, 719 (1979)].
  
\bibitem{Starobinsky:1980te}
  A.~A.~Starobinsky,
  Phys.\ Lett.\  B {\bf 91}, 99 (1980).
  
\bibitem{Rubakov:1982df}
  V.~A.~Rubakov, M.~V.~Sazhin and A.~V.~Veryaskin,
  Phys.\ Lett.\  B {\bf 115} (1982) 189.
  
\bibitem{Spergel:2006hy}
  D.~N.~Spergel {\it et al.}  [WMAP Collaboration],
  Astrophys.\ J.\ Suppl.\  {\bf 170} (2007) 377
  [arXiv:astro-ph/0603449].
   
\bibitem{Veneziano:1991ek}
  G.~Veneziano,
  Phys.\ Lett.\ B {\bf 265} (1991) 287.
  
  
\bibitem{Gasperini:2002bn}
  M.~Gasperini and G.~Veneziano,
  Phys.\ Rept.\  {\bf 373} (2003) 1
  [arXiv:hep-th/0207130].
  
\bibitem{Sakharov:1967dj}
  A.~D.~Sakharov,
  Pisma Zh.\ Eksp.\ Teor.\ Fiz.\  {\bf 5} (1967) 32
  [JETP Lett.\  {\bf 5} (1967) 24].
  
\bibitem{Kuzmin:1985mm}
  V.~A.~Kuzmin, V.~A.~Rubakov and M.~E.~Shaposhnikov,
  Phys.\ Lett.\  B {\bf 155}, 36 (1985).
    
\bibitem{Georgi:1974yf}
  H.~Georgi, H.~R.~Quinn and S.~Weinberg,
  Phys.\ Rev.\ Lett.\  {\bf 33} (1974) 451.
  
\bibitem{Raby:2006sk}
  S.~Raby,
  arXiv:hep-ph/0608183.
   
\bibitem{Dimopoulos:1981yj}
  S.~Dimopoulos, S.~Raby and F.~Wilczek,
  Phys.\ Rev.\  D {\bf 24} (1981) 1681.
  
\bibitem{ArkaniHamed:2004fb}
  N.~Arkani-Hamed and S.~Dimopoulos,
  JHEP {\bf 0506} (2005) 073
  [arXiv:hep-th/0405159].
  
  
\bibitem{Giudice:2004tc}
  G.~F.~Giudice and A.~Romanino,
  Nucl.\ Phys.\  B {\bf 699} (2004) 65
  [Erratum-ibid.\  B {\bf 706} (2005) 65]
  [arXiv:hep-ph/0406088].
  
\bibitem{'tHooft:1971rn}
  G.~'t Hooft,
  Nucl.\ Phys.\  B {\bf 35} (1971) 167.
  
\bibitem{Gildener:1976ai}
  E.~Gildener,
  Phys.\ Rev.\  D {\bf 14} (1976) 1667.
    
\bibitem{'tHooft:1972fi}
  G.~'t Hooft and M.~J.~G.~Veltman,
  Nucl.\ Phys.\  B {\bf 44} (1972) 189.
  
\bibitem{Weinberg:1980wa}
  S.~Weinberg,
  Phys.\ Lett.\  B {\bf 91} (1980) 51.
  
  
\bibitem{Geng:1988pr}
  C.~Q.~Geng and R.~E.~Marshak,
  Phys.\ Rev.\  D {\bf 39} (1989) 693.

\bibitem{Foot:1990uf}
  R.~Foot, G.~C.~Joshi, H.~Lew and R.~R.~Volkas,
  Mod.\ Phys.\ Lett.\  A {\bf 5} (1990) 2721.
   
\bibitem{Georgi:1974sy}
  H.~Georgi and S.~L.~Glashow,
  Phys.\ Rev.\ Lett.\  {\bf 32} (1974) 438.
  
\bibitem{Buras:1977yy}
  A.~J.~Buras, J.~R.~Ellis, M.~K.~Gaillard and D.~V.~Nanopoulos,
  Nucl.\ Phys.\  B {\bf 135} (1978) 66.
  
\bibitem{Hill:1983xh}
  C.~T.~Hill,
  Phys.\ Lett.\  B {\bf 135} (1984) 47.
  
\bibitem{Shafi:1983gz}
  Q.~Shafi and C.~Wetterich,
  Phys.\ Rev.\ Lett.\  {\bf 52} (1984) 875.
   
  
\bibitem{Parida:1989kg}
  M.~K.~Parida, P.~K.~Patra and A.~K.~Mohanty,
  Phys.\ Rev.\  D {\bf 39}, 316 (1989).

\bibitem{Brahmachari:1993yn}
  B.~Brahmachari, U.~Sarkar, K.~Sridhar and P.~K.~Patra,
  Mod.\ Phys.\ Lett.\  A {\bf 8} (1993) 1487.
 
  
\bibitem{Linde:2007fr}
  A.~Linde,
  arXiv:0705.0164 [hep-th].
  
  
\bibitem{Lyth:1998xn}
  D.~H.~Lyth and A.~Riotto,
  Phys.\ Rept.\  {\bf 314} (1999) 1
  [arXiv:hep-ph/9807278].
  
  
\bibitem{Lesgourgues:2007gp}
  J.~Lesgourgues and W.~Valkenburg,
  arXiv:astro-ph/0703625.
  
  
\bibitem{Hwang:1998mx}
  J.~C.~Hwang and H.~Noh,
 Phys.\ Rev.\ Lett.\  {\bf 80} (1998) 4621. 
    [arXiv:astro-ph/9811069].
  
\bibitem{Komatsu:1999mt}
  E.~Komatsu and T.~Futamase,
  Phys.\ Rev.\ D {\bf 59} (1999) 064029. 
  [arXiv:astro-ph/9901127].
  
  
\bibitem{Jackiw:1976pf}
  R.~Jackiw and C.~Rebbi,
  Phys.\ Rev.\ Lett.\  {\bf 37} (1976) 172.
    
\bibitem{Callan:1976je}
  C.~G.~Callan, R.~F.~Dashen and D.~J.~Gross,
  Phys.\ Lett.\  B {\bf 63} (1976) 334.
 
\bibitem{Peccei:1977hh}
  R.~D.~Peccei and H.~R.~Quinn,
  Phys.\ Rev.\ Lett.\  {\bf 38} (1977) 1440.
  
  
\bibitem{Weinberg:1977ma}
  S.~Weinberg,
  Phys.\ Rev.\ Lett.\  {\bf 40} (1978) 223.
  
  
\bibitem{Wilczek:1977pj}
  F.~Wilczek,
  Phys.\ Rev.\ Lett.\  {\bf 40} (1978) 279.
  
\bibitem{Shifman:1979if}
  M.~A.~Shifman, A.~I.~Vainshtein and V.~I.~Zakharov,
  Nucl.\ Phys.\  B {\bf 166} (1980) 493.
  
    
\bibitem{Zhitnitsky:1980tq}
  A.~R.~Zhitnitsky,
  Sov.\ J.\ Nucl.\ Phys.\  {\bf 31} (1980) 260
  [Yad.\ Fiz.\  {\bf 31} (1980) 497].
  
    
\bibitem{Dine:1981rt}
  M.~Dine, W.~Fischler and M.~Srednicki,
  Phys.\ Lett.\  B {\bf 104} (1981) 199.
  
\bibitem{Asztalos:2006kz}
  S.~J.~Asztalos, L.~J.~Rosenberg, K.~van Bibber, P.~Sikivie and K.~Zioutas,
  Ann.\ Rev.\ Nucl.\ Part.\ Sci.\  {\bf 56}, 293 (2006).

 
\bibitem{Raffelt:1990yz}
  G.~G.~Raffelt,
  Phys.\ Rept.\  {\bf 198} (1990) 1.
   
  
\bibitem{Khlebnikov:1987zg}
  S.~Y.~Khlebnikov and M.~E.~Shaposhnikov,
  Phys.\ Lett.\  B {\bf 203} (1988) 121.
  
\bibitem{Khlebnikov:2004am}
  S.~Khlebnikov and M.~Shaposhnikov,
  Phys.\ Rev.\  D {\bf 71} (2005) 104024
  [arXiv:hep-th/0412306].
  
\bibitem{Aldazabal:2002py}
  G.~Aldazabal, L.~E.~Ibanez and A.~M.~Uranga,
  JHEP {\bf 0403} (2004) 065
  [arXiv:hep-ph/0205250].
  
\bibitem{Bars:2006dy}
  I.~Bars,
  Phys.\ Rev.\  D {\bf 74} (2006) 085019
  [arXiv:hep-th/0606045].

\bibitem{Khlebnikov:2006yq}
  S.~Khlebnikov,
  Phys.\ Rev.\  D {\bf 74}, 085007 (2006)
  [arXiv:hep-th/0608070].
  
\bibitem{Seesaw}
P.~Minkowski,
Phys.\ Lett.\ B {\bf 67} (1977) 421;
T.~Yanagida,
Progr.\ Theor.\ Phys.\ {\bf 64} (1980) 1103 ; 
M.~Gell-Mann, P.~Ramond and R.~Slansky, 
in {\em Supergravity}, 
North Holland, Amsterdam 1980.  


\bibitem{Fukugita:1986hr}
  M.~Fukugita and T.~Yanagida,
  Phys.\ Lett.\  B {\bf 174} (1986) 45.
  
\bibitem{Rychkov:2007uq}
  V.~S.~Rychkov and A.~Strumia,
  Phys.\ Rev.\  D {\bf 75} (2007) 075011
  [arXiv:hep-ph/0701104].

\bibitem{Shaposhnikov:1986jp}
  M.~E.~Shaposhnikov,
  JETP Lett.\  {\bf 44} (1986) 465
  [Pisma Zh.\ Eksp.\ Teor.\ Fiz.\  {\bf 44} (1986) 364].
  
\bibitem{Shaposhnikov:1987tw}
  M.~E.~Shaposhnikov,
  Nucl.\ Phys.\  B {\bf 287} (1987) 757.
  
\bibitem{Bertone:2004pz}
  G.~Bertone, D.~Hooper and J.~Silk,
  Phys.\ Rept.\  {\bf 405} (2005) 279
  [arXiv:hep-ph/0404175].
  
\bibitem{deGouvea:2005er}
  A.~de Gouvea,
  Phys.\ Rev.\  {\bf D 72} (2005) 033005.
  [arXiv:hep-ph/0501039].
  
\bibitem{Aguilar:2001ty}
A.~Aguilar {\it et al.}  [LSND Collaboration],
Phys.\ Rev.\ {\bf D 64} (2001) 112007.
[arXiv:hep-ex/0104049].


\bibitem{AguilarArevalo:2007it}
  A.~A.~Aguilar-Arevalo {\it et al.}  [The MiniBooNE Collaboration],
  Phys.\ Rev.\ Lett.\  {\bf 98} (2007) 231801
  [arXiv:0704.1500 [hep-ex]].



\bibitem{Dodelson:1993je}
  S.~Dodelson and L.~M.~Widrow,
  Phys.\ Rev.\ Lett.\  {\bf 72} (1994) 17
  [arXiv:hep-ph/9303287].
  

\bibitem{Moore:1999nt} 
B.~Moore, {\it et al.}, 
  Astrophys.\ J.\  {\bf 524} (1999) L19.
  

\bibitem{Bode:2000gq}
  P.~Bode, J.~P.~Ostriker and N.~Turok,
  Astrophys.\ J.\  {\bf 556} (2001) 93
  [arXiv:astro-ph/0010389].
  
  
\bibitem{Goerdt:2006rw} 
T.~Goerdt, {\it et al.}, 
  Mon.\ Not.\ Roy.\ Astron.\ Soc.\  {\bf 368} (2006) 1073
  [arXiv:astro-ph/0601404].
  
  
\bibitem{Gilmore:2007fy}%
G.~Gilmore, {\it et al.}, 
 Astrophys.\ J.\ {\bf 663} (2007) 948
  arXiv:astro-ph/0703308.
  
   
\bibitem{Penarrubia:2007zz}
  J.~Penarrubia, A.~McConnachie and J.~F.~Navarro,
  arXiv:astro-ph/0701780.
  

   
\bibitem{Sommer-Larsen:1999jx}
  J.~Sommer-Larsen and A.~Dolgov,
  Astrophys.\ J.\  {\bf 551} (2001) 608
  [arXiv:astro-ph/9912166].
  
    
\bibitem{Shi:1998km}
  X.~D.~Shi and G.~M.~Fuller,
  Phys.\ Rev.\ Lett.\  {\bf 82} (1999) 2832
  [arXiv:astro-ph/9810076].

 
\bibitem{Shaposhnikov:2007nf}
  M.~Shaposhnikov,
  arXiv:astro-ph/0703673.
  
  
\bibitem{Ruchayskiy:2007pq}
  O.~Ruchayskiy,
  arXiv:0704.3215 [astro-ph].
  
\bibitem{Boyarsky:2006jm}
  A.~Boyarsky, A.~Neronov, O.~Ruchayskiy and M.~Shaposhnikov,
  JETP Lett.\  {\bf 83}, 133 (2006)
  [arXiv:hep-ph/0601098].
  
\bibitem{Akhmedov:1998qx}
  E.~K.~Akhmedov, V.~A.~Rubakov and A.~Y.~Smirnov,
  Phys.\ Rev.\ Lett.\  {\bf 81}, 1359 (1998)
  [arXiv:hep-ph/9803255].
  
  
\bibitem{Buchmuller:1990pz}
  W.~Buchmuller and C.~Busch,
  Nucl.\ Phys.\ B {\bf 349} (1991) 71.
  
  
   
\bibitem{Coleman:1973jx}
  S.~R.~Coleman and E.~Weinberg,
  Phys.\ Rev.\  D {\bf 7} (1973) 1888.
  
  
\bibitem{Meissner:2006zh}
  K.~A.~Meissner and H.~Nicolai,
  Phys.\ Lett.\  B {\bf 648} (2007) 312
  [arXiv:hep-th/0612165].
  
\bibitem{Bezrukov:2005mx}
  F.~Bezrukov,
  Phys.\ Rev.\  D {\bf 72} (2005) 071303.
  [arXiv:hep-ph/0505247].
  
  
\bibitem{Klapdor-Kleingrothaus:2006ff}
  H.~V.~Klapdor-Kleingrothaus and I.~V.~Krivosheina,
  Mod.\ Phys.\ Lett.\  A {\bf 21} (2006) 1547.
  
  
\bibitem{Boyarsky:2006fg}
  A.~Boyarsky, A.~Neronov, O.~Ruchayskiy, M.~Shaposhnikov and I.~Tkachev,
  Phys.\ Rev.\ Lett.\  {\bf 97} (2006) 261302
  [arXiv:astro-ph/0603660].
  
  
\bibitem{Boyarsky:2006hr}
  A.~Boyarsky, J.~W.~den Herder, A.~Neronov and O.~Ruchayskiy,
  arXiv:astro-ph/0612219.
  

\bibitem{Bezrukov:2006cy}
  F.~Bezrukov and M.~Shaposhnikov,
  Phys.\ Rev.\  D {\bf 75} (2007) 053005
  [arXiv:hep-ph/0611352].


 
 \end{thebibliography}
\end{document}